\documentclass{aa}
\usepackage{times}
\usepackage{graphics}
\begin{document}

\thesaurus{01(08.09.2 HD 93521; 09.01.1; 09.13.2; 13.21.3)}

\title{
{\it Letter to the Editor\/}\\
ORFEUS\,II Echelle spectra:\\
Molecular hydrogen at high velocities toward HD\,93521}
\author{
W.\,Gringel\inst{1} \and
J.\,Barnstedt\inst{1} \and
K.S.\,de Boer\inst{2} \and
M.\,Grewing\inst{3} \and
N.\,Kappelmann\inst{1} \and
P.\,Richter\inst{2,4}}

\offprints{gringel@astro.uni-tuebingen.de}

\institute{Institut f\"{u}r Astronomie und Astrophysik, Abt.
Astronomie, Universit\"{a}t T\"{u}bingen, Waldh\"{a}userstr. 64,
D-72076 T\"{u}bingen, Germany \and
Sternwarte der Universit\"{a}t Bonn, Auf dem H\"{u}gel 71, D-53121
Bonn, Germany \and
Institut de Radio Astronomie Millim\'{e}trique (IRAM), 300 Rue de la
Piscine, F-38406 Saint Martin d'H\`{e}res, France \and
now at Dept. of Astronomy, University of Wisconsin, 475 N. Charter
Street, Madison, Wisconsin 53706, USA}

\titlerunning{Molecular hydrogen at high velocities toward HD\,93521}

\date{Received $<$date$>$ / Accepted $<$date$>$ }

\maketitle

\begin{abstract}

Absorption lines of interstellar molecular hydrogen in the far
ultraviolet (FUV) have been observed in the spectrum of the O9.5 halo
star \object{HD\,93521}, located some 1500\,pc from the Galactic plane.
During the second {\it ORFEUS-SPAS\/} mission a spectrum
with the Echelle spectrometer has been recorded with a total
integration time of 1740\,s. The resolution achieved was about
$\lambda/\Delta\lambda$\,$\geq$\,10.000 with a signal-to-noise ratio of
up to 25. For the first time two components of molecular hydrogen have
been observed in absorption at velocities of
$\simeq$\,$-$12\,km\,s$^{-1}$ in the Galactic disk and at
$\simeq$\,$-$62\,km\,s$^{-1}$ located presumably in the Galactic halo.
The column densities derived from a standard curve of growth analysis
were found to be $N($H$_{2})$\,=\,10$^{17.0}$\,cm$^{-2}$ for the
disk component and $N($H$_{2})$\,=\,10$^{14.6}$ cm$^{-2}$
respectively for the component located in the Galactic halo.

\end{abstract}

\section{Introduction}

Interstellar molecular hydrogen can be investigated in the near
infrared in emission and in the far ultraviolet ({\it FUV\/}) in
absorption. The {\it FUV\/} spectroscopy offers the possibility to
investigate the cool component of the diffuse {\it ISM\/} in which the
H$_{2}$ molecules play a dominant role. The spectral range between 910
and 1150\,{\AA} contains the absorption transitions of the Lyman and
Werner bands of molecular hydrogen. Since the {\it Copernicus\/}
satellite some 20 years ago (Spitzer et al. \cite{spitzer1973},
\cite{spitzer1974a}) no high resolution spectroscopy in the {\it FUV\/}
could be done. A comprehensiv survey of interstellar molecular hydrogen
as observed with the {\it Copernicus\/} satellite is given by Savage et
al. (\cite{savage1977}).

With the {\it ORFEUS-SPAS\,II\/} mission launched aboard the US Space
Shuttle {\it COLUMBIA\/} in Nov.\,1996, it was possible to gather
Echelle spectra with a high resolution of
$\lambda/\Delta\lambda$\,$\geq$\,10.000 of much fainter objects than
observable with {\it Copernicus\/}. The {\it ORFEUS\/} telescope itself
is described in detail by Kr\"{a}mer et al. (\cite{kraemer1988}) and
Grewing et al. (\cite{grewing1991}). An instrument description of the
{\it ORFEUS\,II\/} Echelle spectrometer as well as its performance and
the data reduction are given by Barnstedt et al.
(\cite{barnstedt1999}). Meanwhile the detection of H$_{2}$ absorption
with the {\it ORFEUS\/} Echelle spectrometer was reported for the SMC
by Richter et al. (\cite{richter1998}) as well as for the {\it LMC\/} by de
Boer et al. (\cite{boer1998}). Furthermore molecular hydrogen in the
Galactic halo was observed with {\it ORFEUS \/}by Richter et al.
(\cite{richter1999}) in a high-velocity cloud.

The high latitude Galactic halo star HD\,93521 was one of the PI
targets chosen to be observed for possible H$_{2}$ absorption. The
complete spectrum of HD\,93521 is given by Barnstedt et al.
(\cite{barnstedt2000}). Savage et al. (\cite{savage1977}) reported an
upper limit of $N($H$_{2})$\,$\leq$\,10$^{18.54}$\,cm$^{-2}$ for this
target with a lower limit estimated to be 2.7\,dex smaller. These
authors could not detect other components at higher velocities
presumably because the sensitivity of the {\it Copernicus\/}
spectrometer was too low for such a weak component.

On the other hand it was well known from high resolution ground based
measurements that the interstellar \ion{Ca}{ii}-line shows a complex
profile toward HD\,93521: M\"{u}nch \& Zirin (\cite{muench1961}) found
two main components (out of four altogether) at velocities of
$-$12\,km\,s$^{-1}$ and $-$56.3\,km\,s$^{-1}$ indicating at least one
intermediate velocity cloud ({\it IVC\/}). Later on Rickard
(\cite{rickard1972}) reported a good correlation between the
Ca\,K-lines in front of HD\,93521 to the profile of the hydrogen 21\,cm
lines. Spitzer \& Fitzpatrick (\cite{spitzer1993}) finally deduced from
{\it HST\/} observations with the high resolution {\it GHRS\/} Echelle
spectrometer nine different clouds or filaments toward HD\,93521 for
the less ionized species like \ion{Si}{ii}, \ion{S}{ii} etc. with
heliocentric velocities ranging from $-$66.3\,km\,s$^{-1}$ to
7.3\,km\,s$^{-1}$ and even one more component when fitting radio 21\,cm
observations of neutral hydrogen.

This paper presents the detection of molecular hydrogen in absorption
in the Galactic disk as well as in an {\it IVC\/} in the Galactic halo.
Following the arguments and findings by M\"{u}nch \& Zirin
(\cite{muench1961}), Rickard (\cite{rickard1972}), Albert
(\cite{albert1983}) and Danly et al. (\cite{danly1992}) we designate
absorption features occuring at radial velocities near
$-$12\,km\,s$^{-1}$ as produced by gas components belonging to the
Galactic disk and the absorption features near $-$62\,km\,s$^{-1}$ as
gas belonging to the Galactic halo.

\begin{figure}[tbp]
\centerline{
\resizebox{\hsize}{!}{\includegraphics{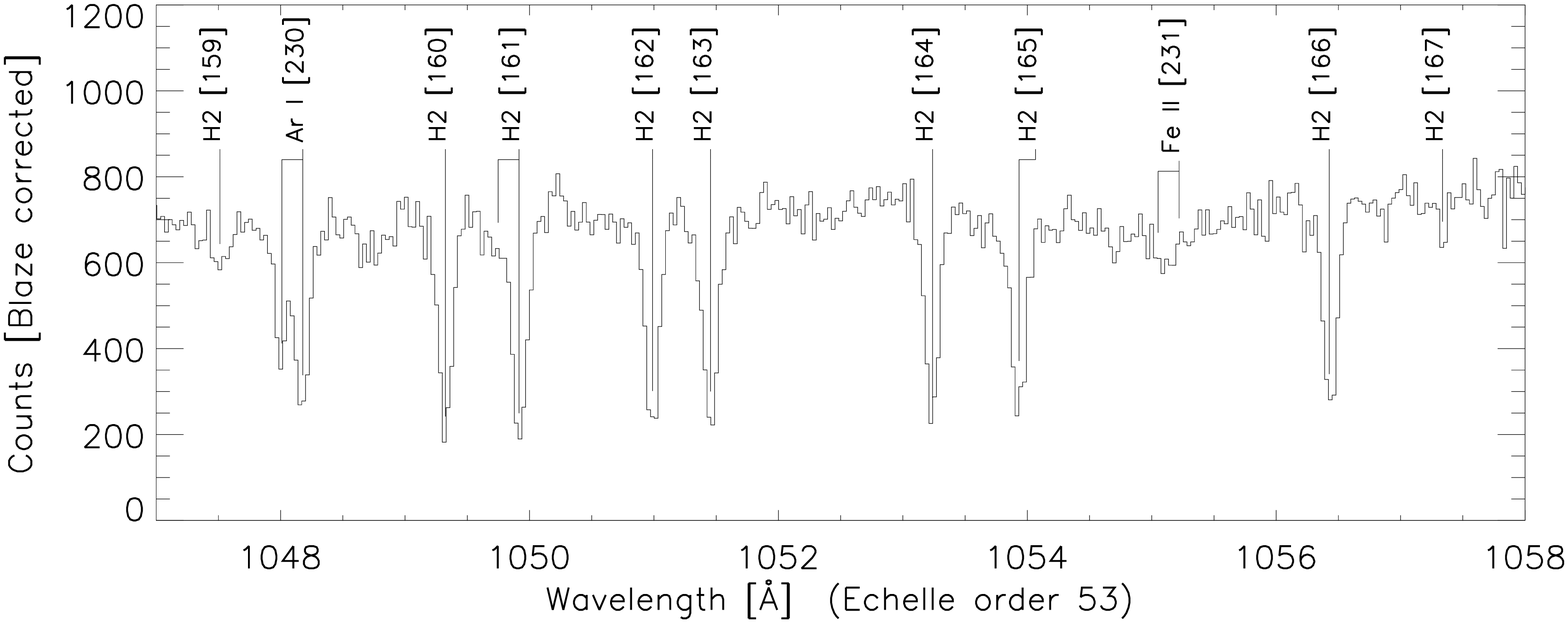}}
}
\caption{A portion of the {\it ORFEUS\/}-Echelle spectrum is shown near
1053\,{\AA} without any smoothing. Besides \ion{Ar}{i} and \ion{Fe}{ii}
the H$_{2}$ absorption lines for the transitions R0 to P3 of the Lyman
series 4-0 (numbers 160 to 166) are very pronounced. The vertical
markings indicate the radial velocities of $-$12\,km\,s$^{-1}$ and
$-$62\,km\,s$^{-1}$ for these lines. The numbers in brackets refer to
the line identifications for HD 93521 given by Barnstedt et al.
(\cite{barnstedt2000})}

\label{fig1}
\end{figure}

\begin{table}[tbp]

\caption[]{H$_{2}$ column densities toward HD\,93521. Galactic disk
component near $-$12\,km s$^{-1}$ (for uncertainties see error bars in
Fig.~\ref{fig3})}

\label{tab1}

\begin{center}
\begin{tabular}{rrrr}

\hline

Rotation & $\log N(J)$ & $b$-value & Number of\\
level $J$ &  & [km s$^{-1}$] & lines used\\
\hline
0 & 16.33 & 7 &   8\\
1 & 16.72 & 7 & 18\\
2 & 16.07 & 7 & 26\\
3 & 15.90 & 7 & 28\\
4 & 14.56 & 7 &   6\\
\hline
\end{tabular}
\end{center}
\end{table}

\section{Observations and data reduction}

HD\,93521 is a high galactic latitude halo star located at
$l$\,=\,183.1\degr, $b$\,=\,62.1\degr, $d$\,=\,1.64\,kpc. The star is
of spectral type O9.5Vp, has $V$\,=\,7.04 mag and
$E(B-V)$=0.02\,mag (Diplas \& Savage, 1994). These authors determined
the total interstellar neutral hydrogen column density from Ly-$\alpha$
absorption gained with the {\it IUE\/} satellite as
$N($\ion{H}{i}$)$\,=\,10$^{20.11}$\,cm$^{-2}$ toward HD\,93521, in
excellent agreement with the sum of the 10 different velocity
components published by Spitzer \& Fitzpatrick (\cite{spitzer1993})
from \ion{H}{i} 21\,cm observations.

The total observing time with the {\it ORFEUS\,II\/} Echelle
spectrometer was 1740\,s in two pointings during two successive orbits.
Both spectra were integrated separately on board and later on coadded
applying the standard extraction procedure described by Barnstedt et
al. (\cite{barnstedt1999}). The data in the different echelle orders
were blaze corrected, after the subtraction of the background caused in
large part by straylight of the echelle grating. Due to the fact that
HD 93521 was not absolutely centered in the entrance diaphragm an
additional radial velocity correction of $-$10\,km\,s$^{-1}$ was
applied to the spectrum (see Barnstedt et al. \cite{barnstedt2000}).

\begin{figure}[tbp]
\centerline{
\resizebox{0.85\hsize}{!}{\includegraphics{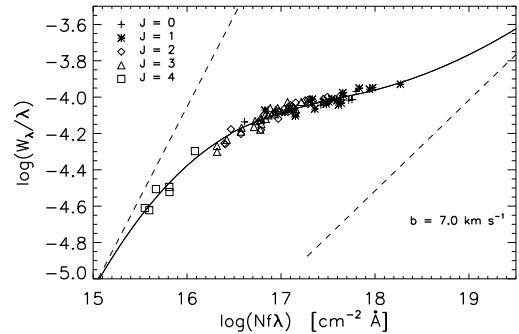}}
}
\caption{The H$_{2}$ lines of the Galactic disk components with radial
velocities $\simeq-12$\,km\,s$^{-1}$ were fitted to a single curve of
growth, indicating $b$\,$\simeq$\,7\,km\,s$^{-1}$. The column
densities for the levels $J$=0--4 are given in
Table~\ref{tab1}. The dashed lines represent the linear part and the
square root section of the curve of growth}

\label{fig2}
\end{figure}

\begin{figure}[tbp]

\centerline{ \resizebox{0.85\hsize}{!}{\includegraphics{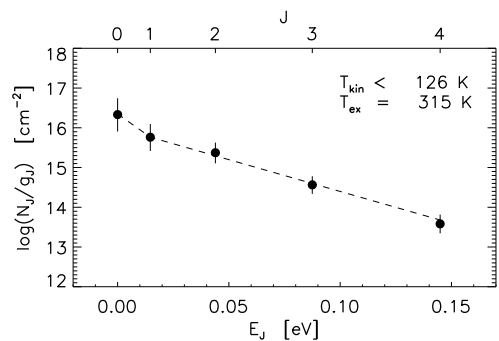}} }
\caption{The H$_{2}$ column densities of the Galactic disk component
divided by their statistical weights are plotted against their
excitation energy for the levels $J$=0--4. The higher levels were
fitted to an equivalent excitation temperature of $\simeq$\,315\,K. For
the two lower levels the kinetic excitation temperature is $<$\,126\,K
(see sect.\,3). The error bars shown are based on the uncertainties in
the curve of growth fits}

\label{fig3}
\end{figure}

\section{H$_{2}$ column density in the Galactic disk component}

A closer inspection of the absorption line profiles from the Echelle
spectrometer reveals immediately the presence of two components with
different radial velocities as shown in Fig.~\ref{fig1} for the
\ion{Ar}{i} $\lambda$\,1048.2 line. The same holds for \ion{N}{i} and
almost all of the less ionized species like \ion{Si}{ii}, \ion{S}{ii}
and \ion{Fe}{ii} (see also Barnstedt et al. \cite{barnstedt2000}). Both
components are separated by $\simeq$\,50\,km\,s$^{-1}$ and show almost
comparable intensities for the atoms or ions mentioned above.

The H$_{2}$ absorption lines of the Lyman (4-0)-band from R0 to P3,
shown in Fig.~\ref{fig1} also, are very sharp and pronounced. The FWHM
values for some H$_{2}$ absorption lines are sometimes as small as
100\,m{\AA}. The equivalent widths $W_{\lambda}$ of the lines were
determined either directly from the observations in the standard manner
or by a Gaussian fit of the line or if possible by both methods. The
$f$-values for the further analysis were taken from Morton \&
Dinerstein (\cite{morton1976}) for the H$_{2}$ transitions and for the
atomic lines from the compilation of Morton (\cite{morton1991}).

Furthermore, curves of growth have been constructed for each of the
absorptions by the 5 rotational states ($J$=0--4) for the
Galactic disk component located at radial velocities around
$-$12\,km\,s$^{-1}$. The sample of the
$\log (W_{\lambda}/\lambda)$-values for each rotational level $J$
has been shifted horizontally to give a fit to a theoretical
single-cloud curve of growth as a function of $N(J)f\lambda$.
Because the curve of growth just begins to depend slightly
on the damping constant $\gamma$ for the three righthand points in
Fig.~\ref{fig2}, $\gamma$ has been chosen to $\gamma$\,=\,12~10$^8$\,s,
a mean value for these three points. The best fit for the 5
rotational states was obtained with $b$\,=\,7\,km\,s$^{-1}$ and
is shown in the empirical curve of growth in Fig.~\ref{fig2}. The
column densities $N(J)$ obtained in this way can be found in
Table~\ref{tab1}. The uncertainties in the column densities are based
on the respective determinations of the equivalent widths as well as on
the quality of the fits to the curve of growth. They range from 0.25 to
0.45\,dex and are shown in Fig.~\ref{fig3}. The total logarithmic
column density for these lowest 5 rotational levels was found to be
$\log N($H$_{2})$\,=\,17.0\,$\pm$0.4 for the Galactic disk component.

In order to get information about the mean excitation temperature in
the Galactic disk components, we fitted the population densities by a
Boltzmann distribution, as shown in Fig.~\ref{fig3}. The column
densities $N(J)$ divided by their statistical weigths $g_J$ are plotted
against the excitation energy $E_J$. For the two lower rotational
states we derive an upper limit for the excitation temperature $T_{0,1}
< 126$ K for the disk gas. This must be an upper limit because the
column density of the $J$=1 level fits very well to a Boltzmann
distribution for the levels $J$=1--4 (see below). Assuming that the
collisional excitation to level $J$=1 amounts to only 10\% of the
observed column density (in view of the good fit for $J$=1--4) one
finds $T_{0,1} \simeq 47$\,K. This value is in the range of excitation
temperatures reported by Savage et al. (\cite{savage1977}) for general
galactic gas.

\begin{figure}[tbp]

\centerline{
\resizebox{0.73\hsize}{!}{\includegraphics{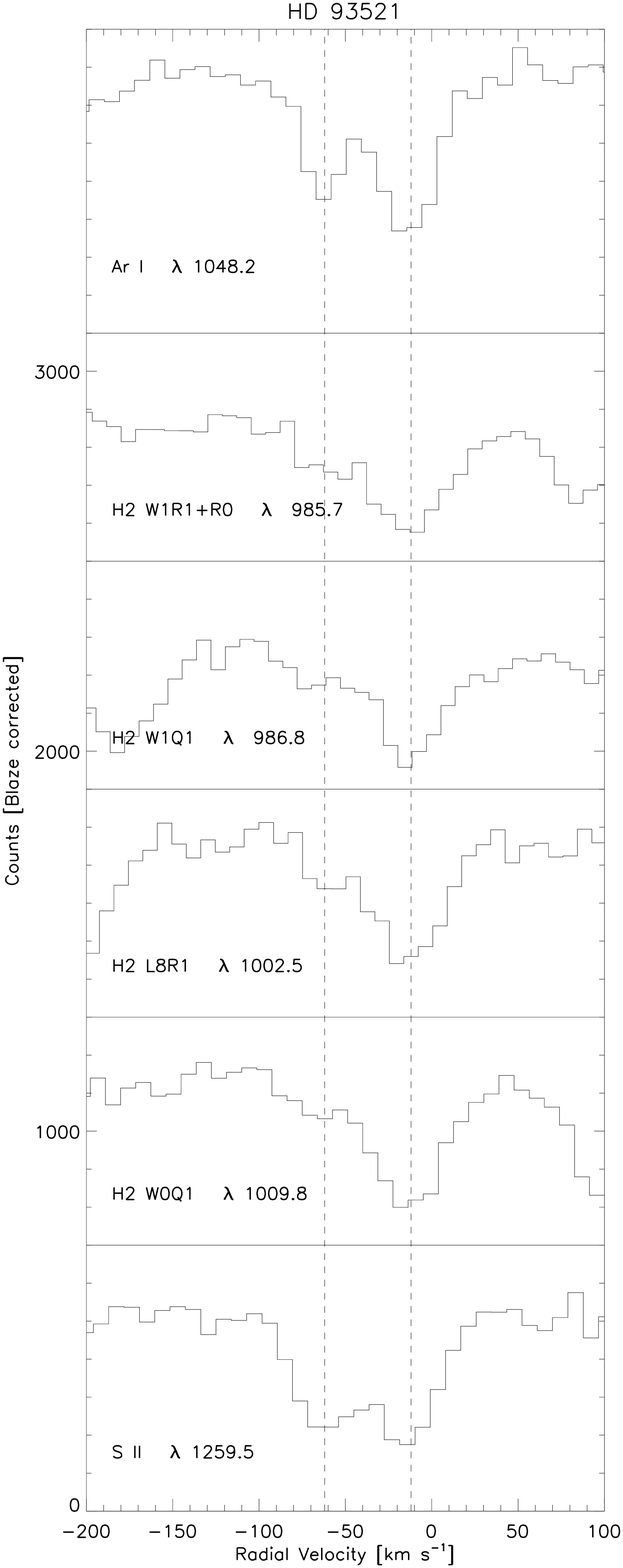}}
}
\caption{Radial velocity plots for \ion{Ar}{i} (top) and \ion{S}{ii}
(bottom) together with the four H$_{2}$ absorption lines shown,
indicate two radial velocity components at $\simeq$\,$-$12 km\,s$^{-1}$
and at $\simeq$\,$-$62\,km\,s$^{-1}$ (vertical dashed lines). The
x-axes of the plots correspond to zero counts for all the lines shown.
The two velocity components are attributed to gas belonging to the
Galactic disk and to the Galactic halo, respectively}

\label{fig4}
\end{figure}

For the rotational levels 1\,$\leq$\,$J$\,$\leq$\,4 we derive a mean
excitation temperature of $\simeq$\,315\,K, indicating that moderate UV
pumping is responsible for the excitation of these states (Spitzer \&
Zweibel \cite{spitzer1974b}).
The results of Fig.~\ref{fig3} indicate furtheron that these ortho
and para levels of H$_{2}$ are in thermal equilibrium in the Galactic
disk gas.

Although a mean excitation temperature of
$\simeq$\,290\,K can be fitted for all rotational levels as indicated
by the data and their uncertainties, we still believe that the bend in
the excitation temperature at the $J$=1 level is real for the
following reasons: The major part in the uncertainty given is due to
the uncertainty of the continuum determination in evaluating the
corresponding equivalent widths. Because the values for $J$=0 and $J$=1
are located on the flat part of the curve of growth, the resulting
absolute errors are the largest ones for these two levels. On the other
hand the R0 and R1 absorption features in the different Lyman-bands are
separated by a small wavelength difference ($<$\,1\,{\AA}) only. Most
of these equivalent widths have been determined with a similar or even
the same value for the adopted continuum. This implies that the
ratio between the column densities for the rotational levels
$J$=0 and $J$=1 should be real and thus that the bend at $J$=1 in
Fig.~\ref{fig3} leading to the two different excitation temperatures
is real, too.

\begin{table}[tbp]

\caption[]{Equivalent widths for rotational level $J$=1. Galactic halo
component near $-$62 km s$^{-1}$ ($^*$\,6\,m{\AA} possible contribution
from L11\,P5 disk gas subtracted)}

\label{tab2}

\begin{center}
\begin{tabular}{lrlr}

\hline

Transition & Wavelength & Equiv. width & $f$-value\\
\,&[{\AA}] & [m{\AA}] &  \\
\hline
L4\,R1 & 1049.958 & 22 & 0.0160\\
L7\,R1 & 1013.434 & 24 & 0.0205\\
W0\,Q1 & 1009.772 & 26 & 0.0238\\
L8\,P1 & 1003.304 & 14 & 0.0084\\
L8\,R1 & 1002.457 & 21 & 0.0181\\
W1\,Q1 &  986.798 & 28$^*$ & 0.0364\\
\hline
\end{tabular}
\end{center}
\end{table}

\section{H$_{2}$ column density in the Galactic halo component}

\begin{figure}[tbp]
\centerline{
\resizebox{0.85\hsize}{!}{\includegraphics{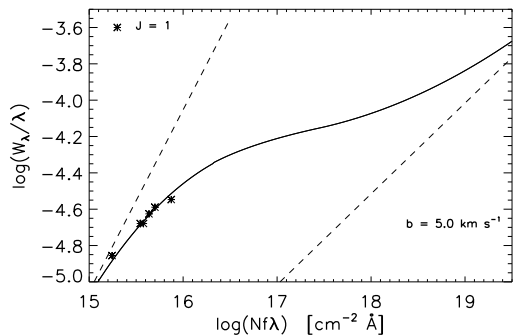}}
}
\caption{H$_{2}$ absorption line components near
$\simeq$\,$-$62\,km\,s$^{-1}$ of the rotational level $J$=1 were
fitted to the indicated curve of growth}

\label{fig5}

\end{figure}

With the {\it ORFEUS\,II\/} Echelle spectrometer it was possible for
the first time to observe H$_{2}$ absorption components at higher
velocities toward HD\,93521. A comparison for some H$_{2}$ absorption
features in the rotational level $J$=1 with \ion{Ar}{i} (see also
Fig.~\ref{fig1}) and a typical \ion{S}{ii} line is shown on a radial
velocity scale in Fig.~\ref{fig4}. These unsmoothed data show
convincingly the presence of a higher velocity component shifted by
about $-$50\,km\,s$^{-1}$ against the Galactic disk component discussed
above. Altogether for six $J$=1 features the corresponding equivalent
widths were determined using a multi-Gaussian fit with linear
continuum. The results are shown in Table~\ref{tab2}.
The {\it IVC\/} component of the W1\,Q1 line in Fig.~\ref{fig4} might be
contaminated by the Galactic disk component of the L11\,P5 line
centered at $-$97\,km\,s$^{-1}$. A maximal contribution of
$W_{\lambda}$\,=\,6\,m{\AA} from the latter has been subtracted from
the W1\,Q1 component. The equivalent widths of Table~\ref{tab2} have
been fitted to a curve of growth as shown in Fig.~\ref{fig5} with a
$b$-value of $\simeq$\,5\,km\,s$^{-1}$. The resulting logarithmic
column density is $\log N(J$=1$)$\,=\,14.32 with an uncertainty of
$\pm$0.25\,dex, the latter arising in large part from the errors made
in determining the equivalent widths.

In the rotational level $J$=2 just one line exhibits an absorption
feature near $-$62\,km\,s$^{-1}$ above our detection limit. The
equivalent width $W_{\lambda}$\,$\leq$\,16\,m{\AA} of this W2\,R2 line
(965.793\,{\AA}, $f$\,=\,0,0323) leads to a value of $\log
N(J$=2$)$\,$\simeq$\,13.88 applying the curve of growth shown in
Fig.~\ref{fig5}. With this value and the population density for
rotational level $J$=1 we calculate an excitation temperature
$T_{1,2}$\,$\simeq$\,800\,K. This value must be regarded as an upper
limit.

The fact that comparable absorption features for the $J$=0 rotational
level in the Lyman-bands have not been observed implies a lower limit
for the corresponding excitation temperature. Assuming again a Boltzman
distribution for the population densities of these rotational levels an
excitation temperature of 300\,K leads to a logarithmic column density
$\log N(J$=0$)$\,$\simeq$\,13.6 and therewith to equivalent widths
below 10\,m{\AA}, and thus below the detection limit of the {\it
ORFEUS\/} Echelle spectrometer.
The stronger R0 components of the Werner
bands could not be used for
this evaluation because they are superimposed by the corresponding
R1 components. In the W1-0 band the wavelength difference of both
components is 19\,m{\AA}, corresponding to about 0.6 electronic pixel.
Within the above excitation temperature
range we estimate the total logarithmic column density of molecular
hydrogen to be $\log N($H$_{2})$\,=\,14.6\,$\pm$0.35 for the {\it
IVC\/} located in the Galactic halo toward HD\,93521.

\section{Concluding remarks}

The {\it ORFEUS FUV\/} spectrum of HD\,93521 shows absorption by
interstellar H$_{2}$ at two radial velocity components around $-$12
km\,s$^{-1}$ and $-$62\,km\,s$^{-1}$.
The hydrogen fraction in its molecular form is 0.0025 for the Galactic
disk gas and about 190 times smaller for the Galactic halo component.
These calculations are based on the \ion{H}{i} column densities
reported by Spitzer \& Fitzpatrick (\cite{spitzer1993}). Attributing
their velocity components 1-4 ($-$66.3 to $-$38.8\,km\,s$^{-1}$) as
\ion{H}{i} gas belonging to the Galactic halo we get $\log
N($\ion{H}{i}$)_{\rm halo}$\,=\,19.69 and $\log
N($\ion{H}{i}$)_{\rm disk}$\,=\,19.88 from the other 6 velocity components.
The estimated range for the excitation temperature indicates that UV
pumping also takes place in the gas of the {\it IVC\/} in the Galactic
halo.

\begin{acknowledgements}

ORFEUS could only be realized with the support of all our German and
American colleagues. The ORFEUS program was supported by DARA grants
WE3\,OS\,8501 and WE2\,QV\,9304 and NASA grant NAG5-696.
We deeply regret the premature passing of our friend and colleague
Gerhard Kr\"{a}mer, the German Project Scientist of the {\it ORFEUS}
project.

\end{acknowledgements}

\end{document}